\documentclass{article}

\PassOptionsToPackage{numbers, square}{natbib}
\usepackage[final]{neurips_2024}

\usepackage[utf8]{inputenc} 
\usepackage[T1]{fontenc}    
\usepackage{hyperref}       
\usepackage{url}            
\usepackage{booktabs}       
\usepackage{amsfonts}       
\usepackage{nicefrac}       
\usepackage{microtype}      
\usepackage{xcolor}         
\usepackage{amsmath}
\usepackage{amssymb}
\usepackage{mathtools}
\usepackage{xspace}
\usepackage{tikz}
\usepackage{xcolor}

\usetikzlibrary{bayesnet}
\DeclarePairedDelimiterX{\infdivx}[2]{(}{)}{%
  #1\;\delimsize\|\;#2%
}
\newcommand{\kl}{D_{KL}\infdivx}
\newcommand{\nullmean}{\mu_{\emptyset}}
\newcommand{\method}{ContrastiveVI+\xspace}

\title{Modeling variable guide efficiency in pooled CRISPR screens with \method}

\author{%
  Ethan Weinberger\thanks{Work performed during an internship at Insitro.} \\
  University of Washington\\
  Seattle, WA 98195 \\
  \texttt{ewein@cs.washington.edu} \\
  \And
  Ryan Conrad \\
  Insitro \\
  South San Francisco, CA 94080 \\
  \texttt{rconrad@insitro.com}
  \And
  Tal Ashuach \\
  Insitro \\
  South San Francisco, CA 94080 \\
  \texttt{tal.ashuach@insitro.com}
}

\begin{document}

\maketitle

\begin{abstract}
Genetic screens mediated via CRISPR-Cas9 combined with high-content readouts have emerged as powerful tools for biological discovery. However, computational analyses of these screens come with additional challenges beyond those found with standard scRNA-seq analyses. For example, perturbation-induced variations of interest may be subtle and masked by other dominant source of variation shared with controls, and variable guide efficiency results in some cells not undergoing genetic perturbation despite expressing a guide RNA. While a number of methods have been developed to address the former problem by explicitly disentangling perturbation-induced variations from those shared with controls, less attention has been paid to the latter problem of noisy perturbation labels. To address this issue, here we propose \method, a generative modeling framework that both disentangles perturbation-induced from non-perturbation-related variations while also inferring whether cells truly underwent genomic edits. Applied to three large-scale Perturb-seq datasets, we find that \method better recovers known perturbation-induced variations compared to previous methods while successfully identifying cells that escaped the functional consequences of guide RNA expression. An open-source implementation of our model is available at \url{https://github.com/insitro/contrastive_vi_plus}.
\end{abstract}

\section{Introduction}

Advances in single-cell genomics technologies have enabled the profiling of molecular modalities across the central dogma at an unprecedented resolution. Moreover, recently developed genetic screening protocols combining CRISPR-Cas9-mediated genome editing with high-content single-cell readouts, such as Perturb-seq \citep{dixit2016perturb}, hold major promise for identifying the genetic bases of functional phenotypes. Such screens are often conducted in a pooled fashion \citep{bock2022high}, in which a CRISPR guide RNA (gRNA) library is introduced in bulk to a cell population. Individual cells subsequently receive different gRNAs corresponding to different gene perturbations, and the specific perturbation induced in a cell can be determined by recovering the barcode sequence corresponding to an individual gRNA.

Despite the enormous potential of high-content genetic screens, computational analyses of these datasets are unfortunately beset by numerous challenges. Beyond confounding technical sources of noise present in scRNA-seq data generally, such as differences in sequencing depth and over-dispersion in RNA counts, analyses of pooled CRISPR datasets present an additional unique set of difficulties. For example, perturbation-induced variations in the data may be relatively subtle compared to those due to other biological processes, such as cell-cycle-related variations or those due to cellular stress responses \citep{papalexi2021characterizing}. Thus, standard single-cell analysis techniques, such as principal component analysis or generative modeling approaches (e.g. scVI \citep{lopez2018deep}) may fail to capture perturbation effects, as these methods prioritize capturing factors with the highest variance across an entire dataset.

To work around this issue, a line of recent work \citep{jones2022contrastive, weinberger2023isolating, abid2019contrastive, severson2019unsupervised} has developed so-called contrastive latent variable models (cLVMs) based on the principle of contrastive analysis \citep{zou2013contrastive}. Such models explicitly disentangle perturbation-induced variations into a set of \textit{salient} latent variables while factors of variation present in both control and perturbed samples are segregated into a second set of \textit{background} variables. While cLVMs have shown promise for analyzing pooled CRISPR datasets \citep{jones2022contrastive, weinberger2023isolating}, their assumed generative processes disagree with the structure of pooled genetic screens in two important ways. First, while pooled screens measure the effects of many perturbations in a single experiment, standard cLVM models assume a single prior distribution over the salient variables for all perturbed samples. Such models thus may fail to discern perturbations with small effect sizes due to shrinkage toward the shared prior. Second, variable guide efficiency results in a subset of cells escaping the effects of perturbation and acting as control cells despite being labeled with a perturbation \citep{papalexi2021characterizing}. Thus, even when restricted to a single perturbation, the assumption of a single unimodal prior leads to nontrivial model misspecification. 

To resolve these issues, here we introduce ContrastiveVI+, an extension of ContrastiveVI, a previously proposed cLVM for scRNA-seq data, that explicitly accounts for the additional structure in pooled genetic screening datasets. The remainder of this work proceeds as follows. In Section \ref{sec:background} we review cLVMs and related work. We then proceed to describe our proposed generative process (Section \ref{sec:generative_process}) and our corresponding inference procedure (Section \ref{sec:inference}). In Section \ref{sec:results} we apply our method to three pooled genetic screening datasets with scRNA-seq readouts, and we find that ContrastiveVI+ learns representations that exhibit better agreement with prior biological knowledge compared to baseline methods while also successfully identifying cells that escaped perturbation.

\section{Background: Contrastive Latent Variable Models}
\label{sec:background}

Recall that the goal of CA is to disentangle novel perturbation-induced factors of variation from those shared with control samples. To formalize this idea, a number of recent works \citep{abid2019contrastive, severson2019unsupervised, weinberger2023isolating, jones2022contrastive} have developed contrastive latent variable models (cLVMs) using the following framework. Letting $x_i$ denote a perturbed sample (e.g.\ a cell infected with a non-control gRNA), we assume that $x_i$ is generated from a random process parameterized by $\theta$ and conditioned on two sets of latent variables
$z_i$ and $t_i$, i.e.,
\begin{equation*}
    x_i \sim p_{\theta}(x_i \mid z_i, t_i)
\end{equation*}
Here $t_i$ denotes a set of salient latent variables capturing novel perturbation-induced variations, while $z_i$ denotes a set of background latent variables capturing variations shared across control and perturbed samples and which are irrelevant our analysis. Without additional constraints, standard latent variable inference procedures are unlikely to naturally disentangle these two sets of latent variables. To address this issue, we leverage our control samples to impose an inductive bias on our model that implicitly encourages disentanglement. In particular, for a control sample $x^{\varnothing}_j$ we assume
\begin{equation*}
    x_j^{\varnothing} \sim p_{\theta}(x_j^{\varnothing} \mid z_j, 0).
\end{equation*}
In words, we assume that control samples are generated from the same process $p_{\theta}$, but with the salient variables fixed at $0$ to represent their absence. Equipped with this inductive bias, we may then apply standard inference techniques with appropriate modifications to learn these disentangled sets of variables, and previous works have leveraged this idea to explore high-content pooled CRISPR screening data with linear cLVMs \citep{jones2022contrastive} and non-linear cLVMs based on deep neural networks \citep{weinberger2023isolating}. Yet, as discussed previously, standard cLVMs may not be ideal for modeling pooled CRISPR screens due to the assumption of a single unimodal prior over the salient latent variables shared across all perturbations. To resolve this issue, in the next section we propose a new generative model that extends the standard cLVM framework with a richer prior over the salient variables to reflect the additional structure present in pooled genetic screening datasets.

\section{The ContrastiveVI+ Probabilistic Model}
\label{sec:generative_process}

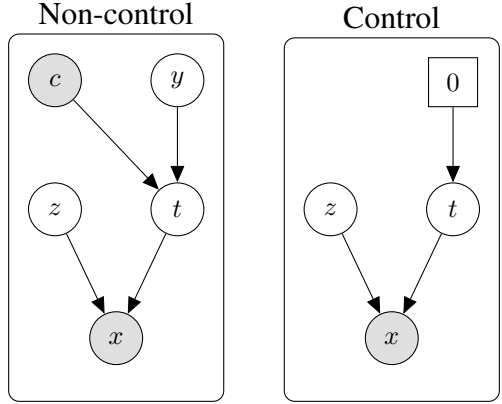
\begin{figure}
\centering
\begin{tikzpicture}
    \node[obs] (x) {$x$};
    \node[latent, above left=of x, xshift=0.4cm, yshift=0.5cm] (z) {$z$};
    \node[latent, above right=of x, xshift=-0.4cm, yshift=0.5cm] (t) {$t$};
    \node[latent, above=of t] (y) {$y$};
    \node[obs, above=of z] (c) {$c$};

    \edge {z} {x};
    \edge {t} {x};
    \edge {y} {t};
    \edge {c} {t};

    \plate[inner sep=0.25cm] {plate_target} {(x)(z)(t)(c)(y)} {};
    \node[above, font=\large] at (current bounding box.north) {Non-control};
\end{tikzpicture}
\qquad
\begin{tikzpicture}[square/.style={regular polygon, regular polygon sides=4}]
    \node[obs] (x) {$x$};
    \node[latent, above left=of x, xshift=0.4cm, yshift=0.5cm] (z) {$z$};
    \node[latent, above right=of x, xshift=-0.4cm, yshift=0.5cm] (t) {$t$};
    \node[square, draw, above=of t] (y) {$0$};

    \edge {z} {x};
    \edge {t} {x};
    \edge {y} {t};

        \plate[inner sep=0.25cm] {plate_target} {(x)(z)(t)(y)} {};
    \node[above, font=\large] at (current bounding box.north) {Control};
\end{tikzpicture}
\caption{Graphical representation of the ContrastiveVI+ generative process for cells with non-control guides (left) and control guides (right).}
\label{fig:graphical_model}
\end{figure}

For a given cell $i$ labelled with a non-control guide RNA $c_i$, let
\begin{equation*}
    z_i \sim \mathcal{N}(0, I_{p})
\end{equation*}
denote a low-dimensional set of latent variables capturing factors of variation found in both perturbed cells as well as controls. Next, let
\begin{equation*}
    y_i \sim \text{Bern}(\alpha)
\end{equation*}
denote a binary value indicating whether a cell expressing a gRNA successfully underwent a corresponding genetic perturbation ($y_i = 1$) or failed to do so ($y_i = 0$). For all of the experiments presented in this work we placed an uninformative prior on $y_i$ with $\alpha = 0.5$. Conditioned on this value and the perturbation label $c_i$, we then draw
\begin{equation*}
    t_i \mid y_i, c_i \sim y_i\cdot\mathcal{N}(\mu_{c}, I_q) + (1 - y_i)\cdot\mathcal{N}(\nullmean, I_{q}).
\end{equation*}
That is, if a cell successfully underwent a genetic perturbation, we assume that its salient latent representation is drawn from a Gaussian centered at a perturbation-specific mean $\mu_c$. For all results presented in this manuscript, we set our perturbation labels $c_i$ as the gene targeted by the guide in cell $i$; however, in principle other labeling schemes to be used (e.g. $c_i$ could denote the specific species of gRNA when multiple gRNAs targeting the same gene are used in an experiment). On the other hand, if the cell was not perturbed, we assume that its salient representation was drawn from a Gaussian with mean $\nullmean$ shared across all perturbations.

Letting $f^{\eta}$ denote a neural network with a softmax non-linearity as the final layer, we then compute
\begin{equation*}
    \rho_{i} = f^{\eta}(z_i, t_i).
\end{equation*}
Analogous to scVI \cite{lopez2018deep}, this vector on the probability simplex represents the expected normalized expression frequency of each gene $g$. For a gene $g$ we then assume that the observed gene expression $x_{ig}$ in cell $i$ is drawn
\begin{equation*}
    x_{ig} \sim \text{ZINB}(\ell_i\rho_{ig}, \theta_{g}, f^{\nu}(z_i, t_i)),
\end{equation*}
where ZINB denotes the zero-inflated negative binomial distribution, $\ell_i$ is the observed library size for cell $i$, $\theta_g$ is a gene-specific inverse dispersion parameter, and $f^{\nu}$ is a neural network whose outputs are interpreted as dropout probabilities.

For a cell $j$ infected with a control gRNA, we assume the same generative process but with $y_j$ fixed at $0$. Thus, cell $j$’s salient variables $t_j$ are always drawn from a Gaussian centered at $\nullmean$, and the region of the salient latent space around $\nullmean$ thus semantically represents the absence of perturbation-induced variations. We depict our generative processes for cells with control and non-control gRNAs in graphical model form in \textbf{Fig. \ref{fig:graphical_model}}.


\section{Inference}
\label{sec:inference}

Exact posterior inference for our model is intractable, so we instead resort to variational inference \citep{blei2017variational} via auto-encoding variational Bayes \citep{kingma2013auto}. For cells with non-control guides, we assume that our variational distribution with parameters $\phi$ factorizes as follows
\begin{equation*}
    q_{\phi}(z_i, t_i, y_i, \mid x_i, c_i) = q_{\phi_z}(z_i \mid x_i)q_{\phi_t}(t_i \mid x_i)q_{\phi_y}(y_i \mid t_i),
\end{equation*}
where $\phi_z$, $\phi_t$, and $\phi_y$ denote parameters of inference networks for $z$, $t$, and $y$ respectively. Here $q(z \mid x)$ and $q(t \mid x)$ take the form of Gaussian distributions, while $q(y \mid t)$ is a Bernoulli distribution. Our corresponding variational bound is then (derivation in Appendix \ref{appendix:elbo}):
\begin{align}
&\mathcal{L}(x_i) =  \mathbb{E}_{q_{\phi_z}(z_i \mid x_i)q_{\phi_t}(t_i \mid x_i)}\left[p_{\theta}(x_i \mid z_i, t_i)\right] - \kl{q_{\phi_z}(z_i \mid x_i)}{p(z_i)} \nonumber \\
&\quad\quad - \mathbb{E}_{q_{\phi_t}(t_i \mid x_i)}\left[\kl{q_{\phi_y}(y_i \mid t_i)}{p(y_i)}\right] \label{eq:pert_elbo} \\
&\quad\quad + \mathbb{E}_{q_{\phi_t}(t_i \mid x_i)}\left[\left(\sum_{y' \in \{0, 1\}}q_{\phi_y}(y' \mid t_i)\left(\log p(t_i \mid y', c_i)\right)\right) - \log q_{\phi_t}(t_i \mid x_i)\right].\nonumber
\end{align}

For cells with non-targeting control (NTC) guides, we assume an alternative variational distribution incorporating our prior knowledge that factorizes as
\begin{equation*}
    q_{\phi_{NTC}}(z_j, t_j, y_j, \mid x_j) = q_{\phi_z}(z_j \mid x_j)\delta\{t_j = \nullmean\}\delta\{y_j = 0\},
\end{equation*}
That is, for control cells we assume that $t_j$ and $y_j$ are fixed at $\nullmean$ and $0$, respectively to reflect the fact that cells with NTC guides are known to be unperturbed ($y_j=0$) and that the salient variables $t_j$ should not capture variations in the observed data for control cells. We also note that the same inference parameters $\phi_z$ are used as in the non-NTC case. We then derive a corresponding bound
\begin{align}
&\mathcal{L}_{NTC}(x^{\varnothing}_j) = \mathbb{E}_{q_{\phi_z}(z_j, \mid x^{\varnothing}_j)}\left[p_{\theta}(x^{\varnothing}_j, | z_j, t_j=\nullmean)\right] - \kl{q_{\phi_z}(z_j \mid x^{\varnothing}_j)}{p(z_j)}.
\label{eq:ntc_elbo}
\end{align}

By fixing $t_j = \nullmean$ for NTC cells during the inference procedure, we ensure that the salient variables $t_j$ do not capture any sources of variation for cells with NTC guides. Thus, as the recognition network parameters $\phi_{z}$ are shared across NTC and non-NTC guide cells, the background variables $z$ are implicitly encouraged to recover sources of variation shared across cells from both groups, while the salient variables $t$ are then free to recover the additional variations only present in perturbed cells. As all cells are assumed to be generated independently, we may then perform inference by maximizing the sums of Equations \ref{eq:pert_elbo} and \ref{eq:ntc_elbo} across all cells via minibatch gradient ascent similar to standard cLVMs \citep{abid2019contrastive, jones2022contrastive, weinberger2023isolating, severson2019unsupervised}. Perturbation-specific means $\mu_c$ along with $\nullmean$ are learned as point estimates and optimized along with our model's other parameters.

While similar implicit schemes have been successfully employed to encourage disentanglement in standard cLVMs, in initial experiments we found that additional regularization was required to ensure that our inference procedure respected the intended semantics of our more structured generative process. First, as the salient space recognition network $q_{\phi_t}$ does not learn from cells with NTC guides when optimizing Equation \ref{eq:ntc_elbo}, we found that $q_{\phi_y}$ did not reliably associate nonperturbed cells to the region of the salient latent space around $\nullmean$. To remedy this issue, we added an additional KL penalty to $\mathcal{L}_{NTC}$ encouraging NTC cells to map to the null region of the salient space, yielding
\begin{align}
&\mathcal{L}^*_{NTC}(x^{\varnothing}_j) = \mathcal{L}_{NTC} - \kl{q(t_j \mid x^{\varnothing}_j)}{\mathcal{N}(\nullmean, I_q)}.
\end{align}
Second, for datasets with larger numbers of perturbations, we observed that perturbation-induced variations were sometimes undesirably captured in the model's background latent space. To discourage this behavior, we leveraged the maximum mean discrepancy (MMD) \cite{gretton2012kernel}, a kernel-based test statistic for determining whether two groups of samples were drawn from the same distribution. In particular, letting $Z^{b}$ denote a minibatch of posterior background latent space samples for cells with non-NTC guides and defining $Z^{b}_{NTC}$ analogously for cells with control guides, we add the penalty
\begin{align}
    -\lambda \cdot \ell_{\text{MMD}}(Z^b, Z^b_{NTC}),
\end{align}
where $\ell_{\text{MMD}}$ denotes the empirical estimate of the MMD of \citet{gretton2012kernel} and $\lambda$ controls the regularization strength. For all experiments presented in this work, we tuned $\lambda$ such that the MMD penalty was of similar magnitude to the KL regularization terms in our ELBOs, a strategy successfully employed in previous work \citep{weinberger2023isolating}. With this penalty, \method’s background latent space was thus explicitly encouraged to capture variations shared across cells with NTC and non-NTC guides.

Combined with our more structured generative process, \method's inference procedure provides two major advantages over that of the original ContrastiveVI model. First, our richer prior on the salient latent space with perturbation-specific means may allow \method to recover subtle perturbation-induced effects in the salient latent space that would be shrunk to the shared unimodal prior in the standard cLVM setup. Second, post-training our approximate posterior $q_{\phi_y}(y_i \mid t_i)$ can be used to classify cells that escaped perturbation versus those that were successfully perturbed.

\begin{figure}
\centering
\includegraphics[width=\textwidth]{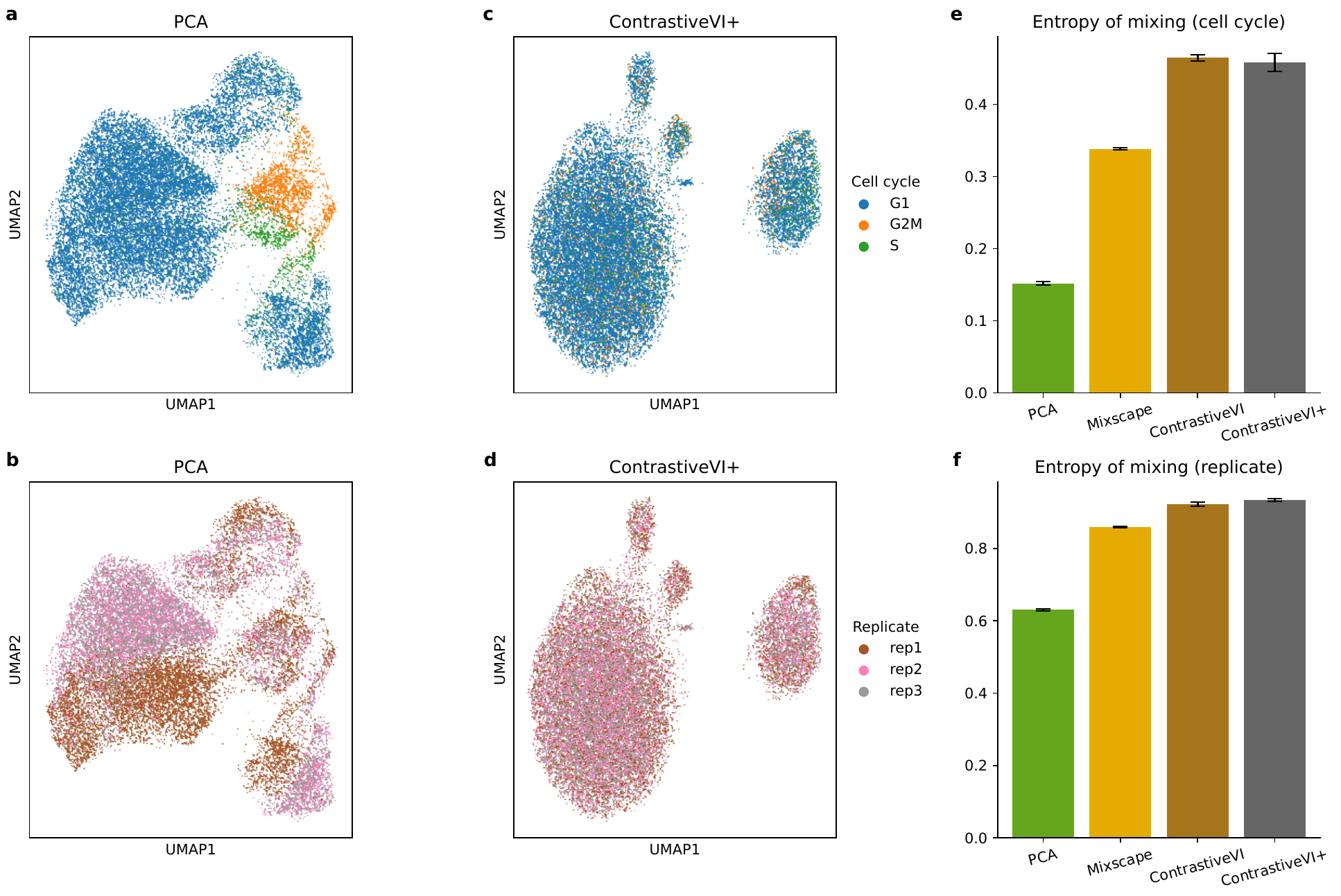}
\caption{\textbf{a-b}, UMAP visualizations of PCA (\textbf{a}) applied to data from \citet{papalexi2021characterizing} colored by cell cycle (\textbf{a}) and replicate (\textbf{b}). \textbf{c-d} UMAP visualizations of \method's salient latent space colored by cell cycle (\textbf{c}) and replicate (\textbf{d}). \textbf{e-f}, Entropy of mixing for \method and baseline methods' representations with respect to cell cycle phase (\textbf{e}) and replicate identity (\textbf{f}).}
\label{fig:papalexi_confounders}
\end{figure}

\section{Experiments}
\label{sec:results}
\textbf{Overview.} To evaluate our method, we applied it to three publicly available pooled genetic screening datasets \citep{papalexi2021characterizing, replogle2022mapping, norman2019exploring}. Based on previous analyses, for each of these datasets we have known confounding sources of variation (e.g.\ cell cycle) and/or known perturbation-induced variations (e.g.\ common gene programs induced by groups of perturbations) that allowed us to assess the quality of \method’s learned representations. Moreover, we also used these datasets to assess the quality of \method’s predictions of escaping versus perturbed cells. Details regarding the preprocessing of these datasets can be found in Appendix \ref{appendix:preprocessing}.

\textbf{Baselines.} For each dataset we benchmarked \method against two previously proposed methods for exploring perturbation-induced variations in pooled genetic screens. First, we considered the original ContrastiveVI model of \citet{weinberger2023isolating} to assess whether our more structured generative process could better recover subtle perturbation-induced variations. Second, we considered the Mixscape method proposed in \citet{papalexi2021characterizing}. Mixscape provides both a procedure for isolating perturbation-induced variations, via a nearest-neighbors-based approach for computing so-called ``perturbation signatures'' for each cell, and a procedure for identifying escaping cells. Finally, as a naive baseline we also include results from simply applying principal component analysis (PCA) to normalized expression levels. Further details on our implementation of \method and baseline methods can be found in Appendix \ref{appendix:baselines}.

\subsection{Initial validation of \method on a small-scale ECCITE-seq dataset}

\begin{figure}
\centering
\includegraphics[width=\textwidth]{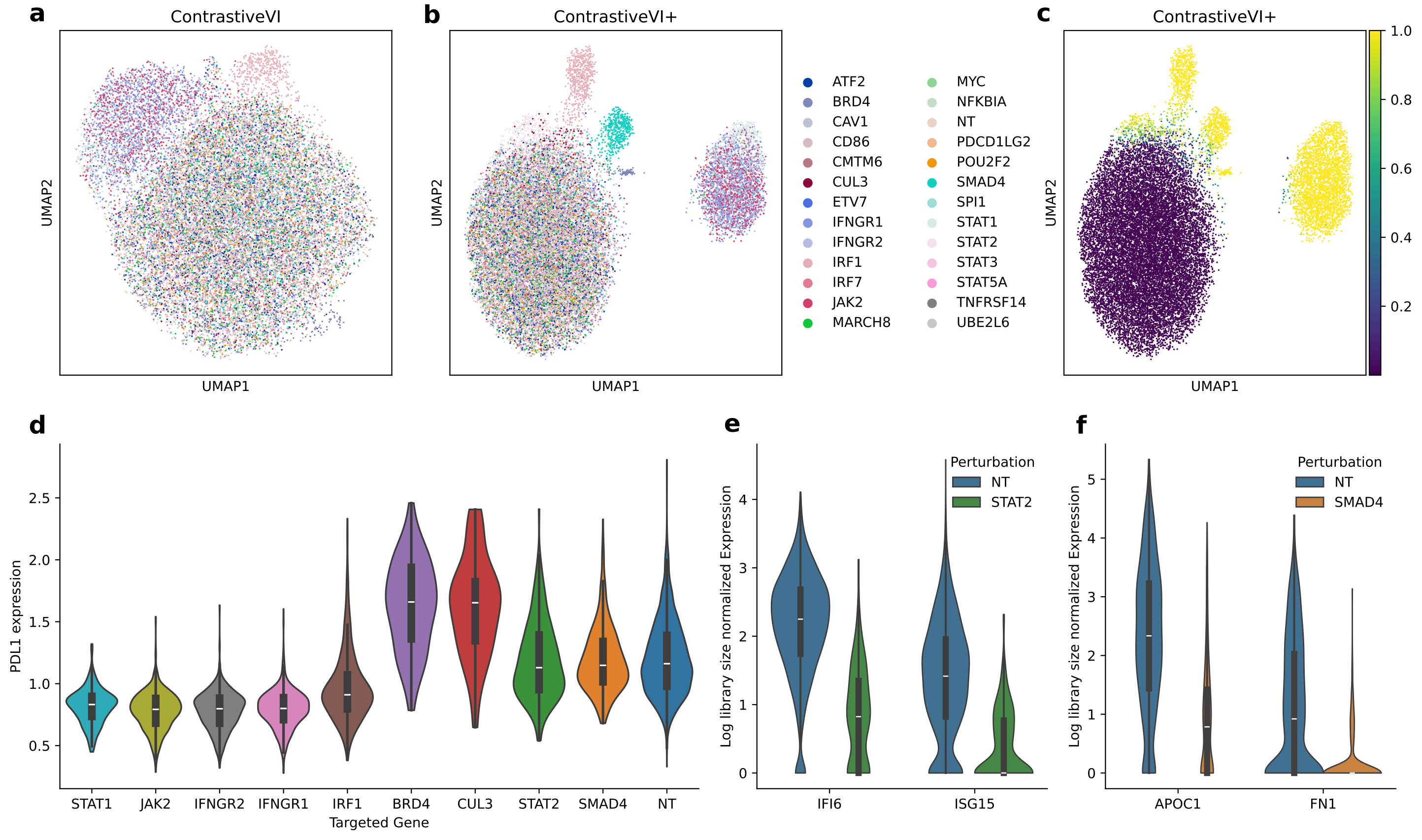}
\caption{\textbf{a-b}, UMAP plots of ContrastiveVI and \method's salient latent representations colored by gene target. \textbf{c}, UMAP plot of \method's salient latent representations colored by inferred probability of perturbation. \textbf{d}, \textit{PDL1} protein expression for gene perturbations highlighted in \method's salient latent space compared to control cells. \textbf{e-f}, Subset of transcriptomic changes for \textit{STAT2}- and \textit{SMAD4}-perturbed cells identified by \method.}
\label{fig:papalexi_structure}
\end{figure}

We first applied \method to data originally presented in \citet{papalexi2021characterizing} collected using ECCITE-seq \citep{mimitou2019multiplexed}, a protocol that combines pooled CRISPR screening with single-cell transcriptomic as well as surface protein measurements. This dataset was collected from a human leukemia monocytic cell line (THP-1) after stimulation with interferon gamma (IFN-$\gamma$), with the goal of identifying immune checkpoint regulators. In this dataset a total of 25 genes were targeted for CRISPR knockout.

Beyond variations due to novel perturbation-induced phenotypes, in their original analyses \citet{papalexi2021characterizing} identified substantial confounding sources of variation shared with control cells due to cell cycle phase and batch effects (\textbf{Fig. \ref{fig:papalexi_confounders}a-b}). We thus began our experiments by assessing \method’s ability to remove these confounding sources of variation in its salient latent space. Qualitatively, we found that \method’s salient latent space was indeed invariant to these confounding variations (\textbf{Fig. \ref{fig:papalexi_confounders}c-d}). To quantify \method and baseline methods’ performance on this task, we computed the entropy of mixing (Appendix \ref{appendix:entropy}) for each method’s representations with respect to cell cycle phase and replicate identity. While all methods resulted in better mixing compared to the naive PCA baseline, we found that ContrastiveVI and \method achieved stronger performance on this task for both confounding sources of variation (\textbf{Fig. \ref{fig:papalexi_confounders}e-f}) compared to Mixscape’s perturbation signatures. This result suggests that the nearest-neighbors based approach employed by Mixscape is less effective at isolating perturbation-induced variations compared to deep generative models.

We next investigated whether \method’s richer model could highlight further trends compared to ContrastiveVI. We found that both methods (\textbf{Fig. \ref{fig:papalexi_structure}a-b}) highlighted a cluster of cells with gRNA’s corresponding to known upstream components of the IFN-$\gamma$ pathway (\textit{IFNGR1}, \textit{IFNGR2}, \textit{JAK2} and \textit{STAT1}), a cluster of cells with gRNAs corresponding to the downtream IFN-$\gamma$ mediator \textit{IRF1}, and a third cluster containing containing cells from all perturbations (including non-targeting gRNAs). Beyond these clusters, \method additionally highlighted clusters of cells expressing gRNAs targeting \textit{SMAD4}, \textit{BRD4}, and \textit{STAT2}. Moreover, when inspecting \method’s inferred values of $y$ (\textbf{Fig. \ref{fig:papalexi_structure}c}), we found that cells in mixed cluster shared with controls were assigned as non-perturbed ($y \approx 0$) while cells in the non-control clusters were classified as perturbed ($y \approx 1$), suggesting that \method successfully distinguished perturbed versus escaping cells.

We verified that these clusterings corresponded to meaningful perturbation effects by first inspecting \textit{PDL1} surface protein expression levels (\textbf{Fig. \ref{fig:papalexi_structure}d}) for cells predicted by \method as perturbed. For most genes, we found corresponding decreases (\textit{IFNGR1}, \textit{IFNGR2}, \textit{JAK2}, \textit{STAT1}, and \textit{IRF1}) or increases (\textit{BRD4}, \textit{CUL3}) in \textit{PDL1} expression compared to control cells. Moreover, while \textit{STAT2} and \textit{SMAD4} did not affect \textit{PDL1} expression, we nevertheless found clear transcriptomic changes in cells predicted by \method as perturbed compared to controls (\textbf{Fig. \ref{fig:papalexi_structure}e-f}). In particlar, \textit{STAT2}-perturbed cells exhibited strong downregulation of interferon-induced genes (e.g.\ \textit{IFI6}, \textit{ISG15}) while \textit{SMAD4}-perturbed cells demonstrated downregulation in inflammatory response genes (e.g.\ \textit{APOC1}, \textit{FN1}).

Taken together, these results illustrate that \method indeed may highlight additional structure in the model’s salient latent space compared to standard cLVMs. Moreover, these results demonstrate that our inference procedure can identify cells exhibiting perturbation effects versus escaping cells.

\subsection{Further validation on a larger-scale CRISPRi screen}

\begin{figure}
\centering
\includegraphics[width=\textwidth]{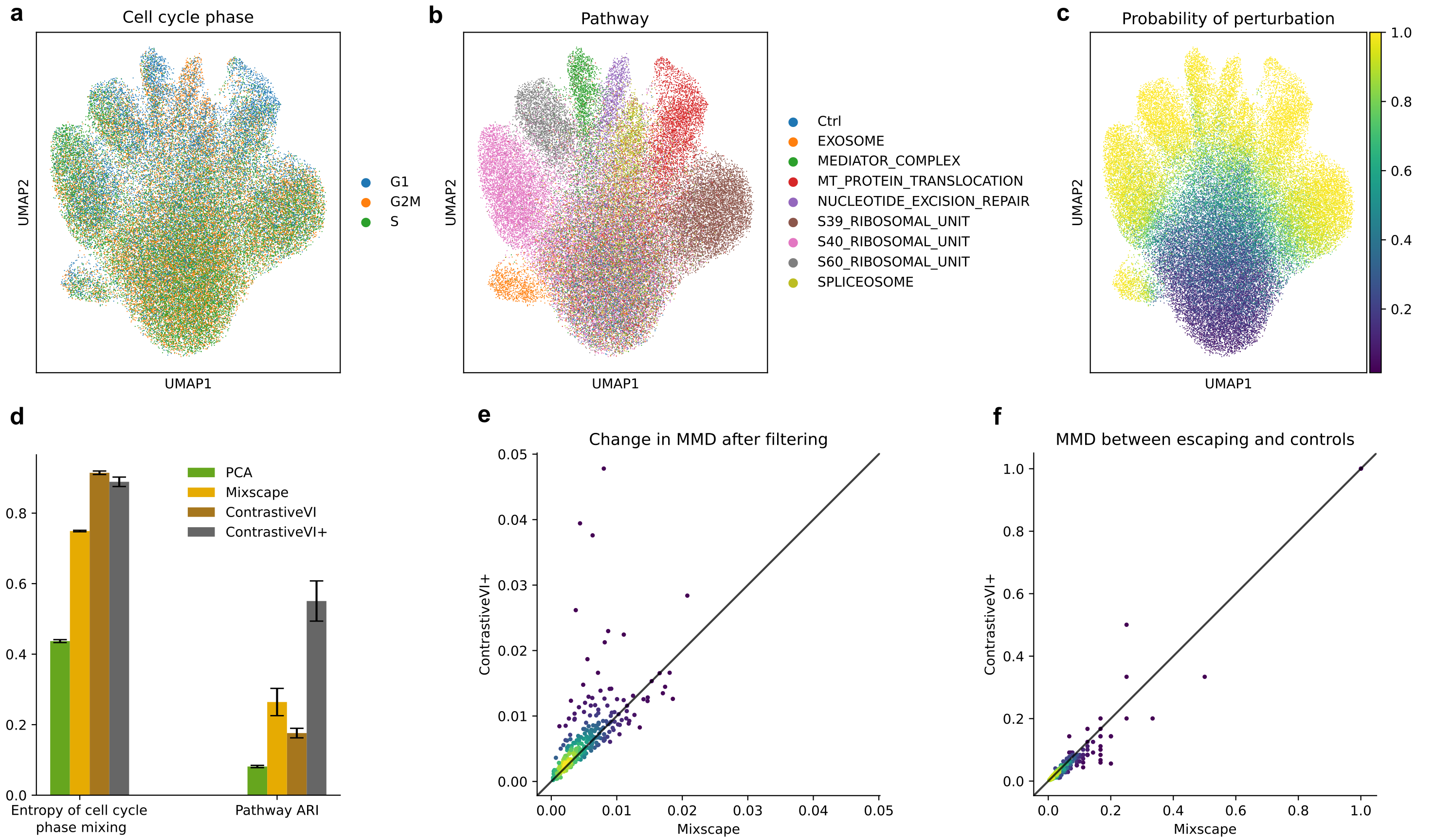}
\caption{\textbf{a-c}, UMAP plots of \method's salient latent space for data from \citet{replogle2022mapping} colored by cell cycle phase (\textbf{a}), pathway annotations (\textbf{b}) and inferred probability of perturbation (\textbf{c}). \textbf{d}, Quantitative assessments of invariance with respect to cell cycle phase (entropy of cell cycle phase mixing) and capturing of known perturbation-induced variations (pathway ARI) for \method and baseline method's salient representations. \textbf{e}, Change in MMD between cells labeled with gRNAs targeting a given gene versus cells with non-targeting control guides after filtering with \method ($y$-axis) or Mixscape ($x$-axis) compared to the MMD without filtering. \textbf{f}, MMD between cells labeled as escaping by \method ($y$-axis) or Mixscape ($x$-axis) versus control cells.}
\label{fig:replogle_latent}
\end{figure}

We next applied ContrastiveVI+ to analyze data from a CRISPR interference (CRISPRi) Perturb-seq dataset presented in \citet{replogle2022mapping}. Following previous work \citep{bereket2024modelling, lopez2023learning}, in our experiments we considered a subset of perturbations identified as having nontrivial effect sizes and which were labeled by the original authors of \citet{replogle2022mapping} as affecting specific biological pathways. In total, we retained data from cells with guides targeting 336 genes as well as cells with NTC guides.

We began our analysis by assessing the quality of ContrastiveVI+’s salient representations. Previous analyses of this data \citep{tu2024supervised} have identified cell cycle as a major confounding source of variation in this dataset shared with control cells. Thus, we would expect \method’s salient latent space to be invariant with respect to cell cycle phase. Moreover, we would expect cells to cluster based on the pathway labels assigned to perturbations by \citep{replogle2022mapping}. We found that \method’s salient space was indeed invariant to cell cycle (\textbf{Fig. \ref{fig:replogle_latent}a}), with clear clusters separating by pathway label (\textbf{Fig. \ref{fig:replogle_latent}b}). Moreover, cells from these distinct pathway clusters were inferred as truly perturbed, while cells in the remaining cluster mixed across pathways and control cells were inferred as escaping (\textbf{Fig. \ref{fig:replogle_latent}c}).

To compare \method and baselines’ performance on this task, we employed the entropy of mixing to measure invariance with respect to cell cycle and used the adjusted rand index (ARI; Appendix \ref{appendix:ari}) to quantify separation based on pathway labels. When computing pathway ARI, for \method and Mixscape we restricted our attention to cells labeled by these methods as truly perturbed to mitigate the impact of escaping cells; for ContrastiveVI we used all cells as this method does not predict perturbed versus escaping cells. We found that \method and ContrastiveVI
had the strongest performance on cell cycle mixing, while \method achieved substantially stronger performance on the pathway ARI metric compared to baselines. These results further demonstrate \method’s superior ability to isolate perturbation-induced variations.

Given its large number of distinct perturbations, we further used this dataset to benchmark \method and Mixscape’s procedures for identifying perturbed versus escaping cells. Intuitively, cells classified as perturbed should have substantially different gene expression profiles compared to cells with NTC guides. To capture this idea, we used the MMD to measure the distance between populations of cells. Specifically, for each gene perturbation we computed the MMD between cells labeled with a gRNA targeting that gene versus cells with NTC guides. We then recalculated the MMD after filtering to cells labeled by \method or Mixscape as perturbed, and finally computed the change in MMD with filtering versus without filtering. We present our results for this experiment in \textbf{Fig. \ref{fig:replogle_latent}e}. Here a \textit{higher} change in MMD indicates that cells classified as perturbed exhibit stronger differences from control cells and thus represents better performance. We found that for a majority of genes \method led to larger changes in MMD compared to
Mixscape, with statistical significance confirmed with a binomial test assuming a null hypothesis of equal chance of either method achieving better performance for each gene ($p < 1\cdot 10^{-5}$).

In isolation, such a metric could be maximized by only retaining cells with the most extreme changes compared to controls and erroneously labeling many truly-perturbed cells as escaping. To counteract this potential pathology, we thus also assessed whether cells labeled as escaping perturbation by each method were indeed similar to cells with NTC guides. To do so, we computed the MMD between cells labeled as escaping by each method versus cells with NTC guides, and we present our results in \textbf{Fig. \ref{fig:replogle_latent}f}. Here lower MMD values indicate that the cells flagged by a method as escaping are closer to true controls and thus represent better performance. We found that cells predicted as escaping by
\method largely had lower MMDs compared to Mixscape ($p < 1\cdot 10^{-5}$, binomial test).

Taken together, these results suggest ContrastiveVI+’s more expressive modeling procedure facilitates superior prediction of perturbed versus escaping cells compared to Mixscape.

\subsection{Exploring the diversity in perturbation responses in a CRISPRa screen}

\begin{figure}
\centering
\includegraphics[width=0.95\textwidth]{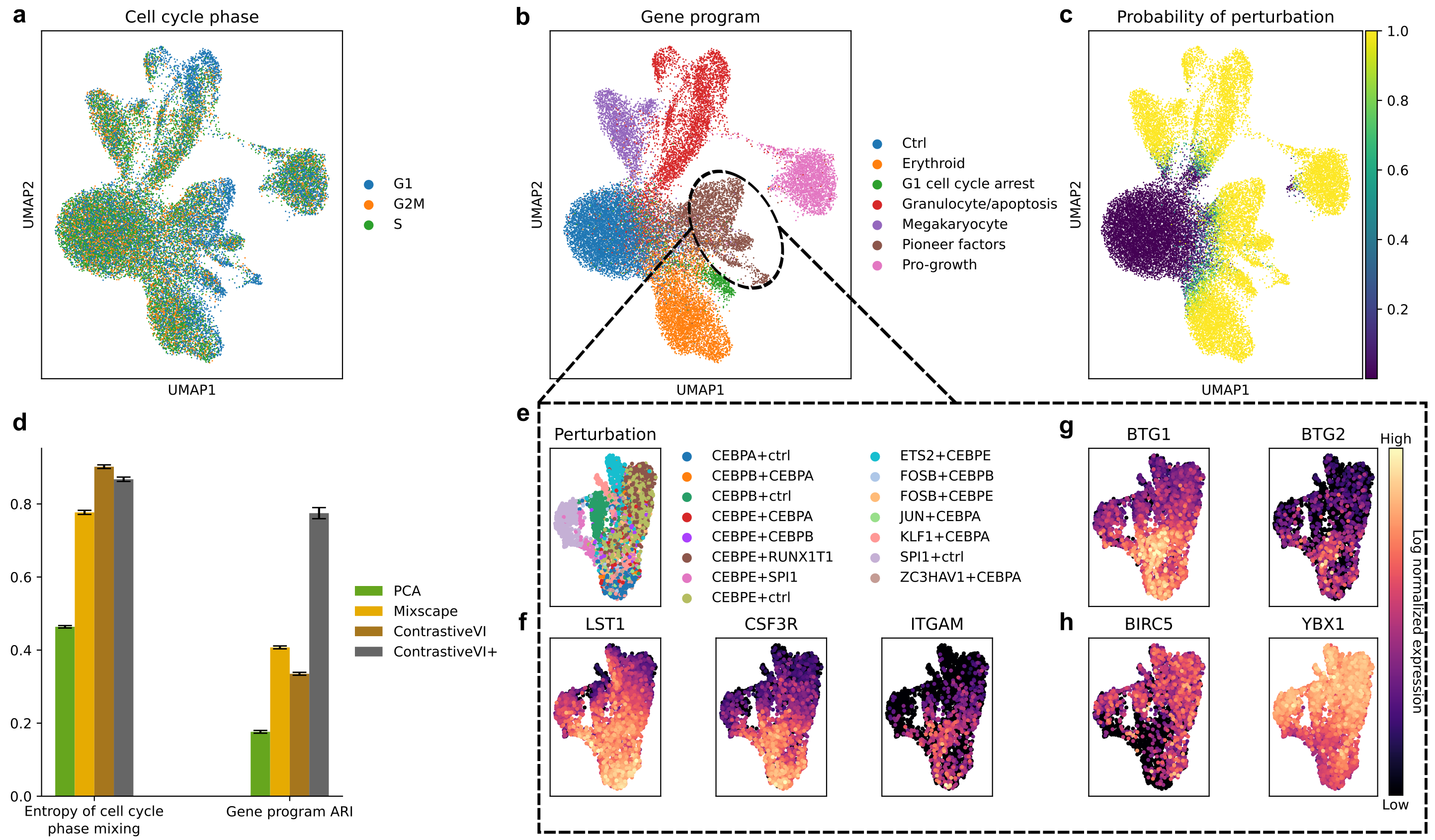}
\caption{\textbf{a-c}, UMAP visualizations of \method's salient latent space for \citet{norman2019exploring} colored by cell cycle phase (\textbf{a}), gene program labels provided by \citet{norman2019exploring} (\textbf{b}), and inferred probability of perturbation (\textbf{c}). \textbf{d}, Quantitative assessments of \method and baseline method's salient representations. \textbf{e-h}, \method's salient latent representations for cells with perturbations labeled as ``granulocyte/apoptosis'' by \citet{norman2019exploring}. Plots colored by perturbation labels (\textbf{e}), canonical granulocyte marker genes (\textbf{f}), the pro-apoptotic anti-proliferation factors \textit{BTG1} and \textit{BTG2} (\textbf{g}), and anti-apoptotic genes \textit{BIRC5} (survivin) and \textit{YBX1} (\textbf{h}).}
\label{fig:norman}
\end{figure}

As a final demonstration of \method’s capabilities, we applied it to explore a Perturb-seq dataset from \citet{norman2019exploring}. In this dataset the authors assessed the effects of CRISPR activation (CRISPRa) perturbations on K562 cells. For our analysis, we focused on a subset of these perturbations labeled in \citet{norman2019exploring} as inducing specific gene programs.

We began by confirming that \method successfully isolated perturbation-induced variations in its salient latent space. Based on the analysis of \citet{norman2019exploring}, we would expect cells to separate by gene program labels. Moreover, we would expect cells to mix across cell cycle phase, a known confounding source of variation shared with control cells in this dataset \citep{weinberger2023isolating}. We found that cells indeed mixed across cell cycle phase (\textbf{Fig. \ref{fig:norman}a}), while separating by gene program labels (\textbf{Fig. \ref{fig:norman}b}), with cells in the separated gene program clusters being predicted by \method as perturbed (\textbf{Fig. \ref{fig:norman}c}). Moreover, \method achieved significantly better separation of gene programs compared to baseline methods as measured by ARI while also achieving strong performance at removing cell cycle effects as measured by entropy of mixing (\textbf{Fig. \ref{fig:norman}{d}}).

In addition to separation between the gene program clusters, in our analysis we also observed separation between substructures within these clusters. To highlight \method's potential to facilitate additional insights, we thus  further inspected the cells with perturbations labeled as ``granulocyte/apoptosis'' and which were predicted by \method as perturbed (\textbf{Fig. \ref{fig:norman}e}). Notably, in \citet{norman2019exploring}, the authors of that work largely analyzed the relationships between perturbations at the pseudobulk level (i.e., by considering the mean expression profile for each perturbation). Thus, a particular focus of our analysis was to see if our single-cell-level model could uncover further relationships between perturbations beyond those discussed in \citet{norman2019exploring}.

First, to understand the genes driving separation in \method's latent space, we employed Hotspot \citep{detomaso2021hotspot}, a tool for identifying informative genes in single-cell data by ranking genes in terms of spatial autocorrelation with respect to a given metric of cell–cell similarity (e.g., the latent space of a VAE). From the top genes returned by Hotspot, we found that separation in \method's salient space was strongly correlated with canonical granulocyte marker genes, such as \textit{LST1}, \textit{CSF3R}, and \textit{ITGAM} (\textbf{Fig. \ref{fig:norman}f}). Moreover, expression of the granulocyte markers was correlated with expression of the pro-apoptotic anti-proliferation factors \textit{BTG1} and \textit{BTG2} (\textbf{Fig. \ref{fig:norman}g}), and inversely correlated with the anti-apoptotic genes \textit{BIRC5}, also known as survivin, and \textit{YBX1} (\textbf{Fig. \ref{fig:norman}h}).

Furthermore, we observed clear heterogeneity in the responses induced by individual perturbations. For example, while some perturbations induced consistently strong upregulation of granulocyte markers (e.g.\ \textit{CEBPA}+ctrl), other perturbations (e.g.\ \textit{CEBPE+ctrl}) resulted in more variable responses. Notably, we often observed strong mixing between cells perturbed solely to activate \textit{CEBPE} (i.e., \textit{CEBPE}+ctrl) and cells perturbed to activate \textit{CEBPE} along with a second gene (e.g. \textit{CEBPE}+\textit{RUNX1T1}, \textit{CEBPE}+\textit{CEBPA}, and \textit{FOSB}+\textit{CEBPE}). This phenomenon suggests that activation of \textit{CEBPE} is sufficient to achieve a certain cellular state, with other perturbations not having an observable impact. Moreover, this behavior is consistent with \textit{CEBPE}'s known function of strongly driving terminal differentiation for granulocytes \citep{theilgaard2022transcription}; in other words, due to \textit{CEBPE} activation inducing cells to differentiate into granulocytes, the effects of additional perturbations may be muted.

Notably, these phenomena were not discussed in \citet{norman2019exploring}, and these results illustrate how the higher resolution of our single-cell-level modeling approach may facilitate insights into the diversity of perturbation responses beyond those possible from previous workflows.

\section{Conclusion}
\label{sec:conclusion}

Here we introduced \method, a deep generative modeling framework for exploring perturbation-induced variations in pooled genetic screening datasets while explicitly accounting for variable guide efficiency and the diversity of responses induced by different perturbations. In experiments on three datasets with scRNA-seq readouts, we found that our model's additional structure resulted in substantially better recovery of known biological relationships compared to baseline methods while also successfully predicting truly perturbed versus escaping cells. Moreover, we found that our more structured modeling approach could reveal further biological insights beyond those provided by other analysis workflows. Future work will involve assessing \method's abilities on larger scale datasets and additional high-content screening modalities beyond scRNA-seq.

\section*{Acknowledgements and Disclosure of Fundingws}

We thank Angela Pisco, Shahin Mohammadi, Srinivasan Sivanandan, and Shreyas Ravishankar for discussions throughout the duration of this project which greatly improved this work. We also thank members of the Data Science and Machine Learning department at Insitro for providing constructive feedback on the results presented in this work.

Disclosures: This work was performed while Ethan Weinberger was an intern at Insitro. Ryan Conrad and Tal Ashuach are employees of Insitro. 

\bibliographystyle{unsrtnat}
\bibliography{main}

\clearpage

\appendix

\section{Derivation of variational bounds for \method}
\label{appendix:elbo}

Here we present a full derivation of the variational bound for cells with non-NTC guides presented in Section \ref{sec:inference}. We begin by assuming that our variational distribution factorizes as 
\begin{equation*}
    q_{\phi}(z_i, t_i, y_i, \mid x_i, c_i) = q_{\phi_z}(z_i \mid x_i)q_{\phi_t}(t_i \mid x_i)q_{\phi_y}(y_i \mid t_i),
\end{equation*}
where $\phi_z$, $\phi_t$, and $\phi_y$ denote parameters of inference networks for $z$, $t$, and $y$ respectively. Leveraging our variational distribution as well as the generative process proposed in Section \ref{sec:generative_process}, we then proceed to derive a corresponding ELBO:

\begin{align*}
&\mathcal{L}(x_i) = \mathbb{E}_{q_{\phi}(z_i, t_i, y_i \mid x_i)}\left[\log\frac{p(z_i, t_i, y_i, x_i \mid c_i)}{q_{\phi}(z_i, t_i, y_i \mid x_i)}\right] \\
&= \mathbb{E}_{q_{\phi_z}(z_i \mid x_i)q_{\phi_t}(t_i \mid x_i)q_{\phi_y}(y_i \mid t_i)}\left[\log \frac{p(z_i)p(t_i \mid y_i, c_i)p(y_i)p(x_i \mid z_i, t_i)}{q_{\phi_z}(z_i \mid x_i)q_{\phi_t}(t_i \mid x_i)q_{\phi_y}(y_i \mid t_i)}\right] \\
&= \mathbb{E}_{q_{\phi_z}(z_i \mid x_i)q_{\phi_t}(t_i \mid x_i)q_{\phi_y}(y_i \mid t_i)}\left[\log \frac{p(z_i)}{q_{\phi_z}(z_i \mid x_i)} + \log \frac{p(t_i \mid y_i, c_i)}{q_{\phi_t}(t_i \mid x_i)} + \log \frac{p(y_i)}{q_{\phi_y}(y_i \mid t_i)} + \right. \\
&\qquad \left. \log p(x_i \mid z_i, t_i)\right] \\
&= \mathbb{E}_{q_{\phi_z}(z_i \mid x_i)q_{\phi_t}(t_i \mid x_i)}\left[p(x_i \mid z_i, t_i)\right]
+ \mathbb{E}_{q_{\phi_z}(z_i \mid x_i)}\left[\log \frac{p(z_i)}{q_{\phi_z}(z_i \mid x_i)}\right] + \\
&\qquad \mathbb{E}_{q_{\phi_t}(t_i \mid x_i)q_{\phi_y}(y_i \mid t_i)}\left[\log \frac{p(t_i \mid y_i, c_i)}{q_{\phi_t}(t_i \mid x_i)}\right] + \mathbb{E}_{q_{\phi_t}(t_i \mid x_i)q_{\phi_y}(y_i \mid t_i)}\left[\log\frac{p(y_i)}{q_{\phi_y}(y_i \mid t_i)}\right]\\
&=  \mathbb{E}_{q_{\phi_z}(z_i \mid x_i)q_{\phi_t}(t_i \mid x_i)}\left[p(x_i \mid z_i, t_i)\right] - \kl{q_{\phi_z}(z_i \mid x_i)}{p(z_i)} \\
&\quad\quad - \mathbb{E}_{q_{\phi_t}(t_i \mid x_i)}\left[\kl{q_{\phi_y}(y_i \mid t_i)}{p(y_i)}\right] \\
&\quad\quad + \mathbb{E}_{q_{\phi_t}(t_i \mid x_i)}\left[\left(\sum_{y' \in \{0, 1\}}q_{\phi_y}(y' \mid t_i)\left(\log p(t_i \mid y', c_i)\right)\right) - \log q_{\phi_t}(t_i \mid x_i)\right]
\end{align*}

For control cells infected with non-targeting control (NTC) guides we assume the following variational distribution:
\begin{equation*}
    q_{\phi_{NTC}}(z_j, t_j, y_j, \mid x^{\varnothing}_j) = q_{\phi_z}(z_j \mid x^{\varnothing}_j)\delta\{t_j = \nullmean\}\delta\{y_j = 0\}.
\end{equation*}

Our corresponding ELBO is then:
\begin{align*}
&\mathcal{L}_{NTC}(x^{\varnothing}_j) = \mathbb{E}_{q(z_j, \mid x^{\varnothing}_j)}\left[\log\frac{p(z_j, x_j \mid t_j=\nullmean, y_j=0)}{q(z_j \mid x^{\varnothing}_j)}\right] \\
&= \mathbb{E}_{q(z_j, \mid x^{\varnothing}_j)}\left[\log\frac{p(x^{\varnothing}_j, | z_j, t_j=\nullmean)p(z_j)}{q(z_j \mid x^{\varnothing}_j)}\right] \\
&= \mathbb{E}_{q(z_j, \mid x^{\varnothing}_j)}\left[p(x^{\varnothing}_j, | z_j, t_j=\nullmean)p(z_j)\right] - \kl{q(z_j \mid x^{\varnothing}_j)}{p(z_j)}.
\end{align*}

\section{Dataset preprocessing}
\label{appendix:preprocessing}

Here we provide descriptions of any preprocessing steps for the datasets considered in this work.

\textbf{\citet{papalexi2021characterizing}.} The cell by gene count matrix along with corresponding metadata for this dataset was obtained from the NIH gene expression omnibus entry \href{https://www.ncbi.nlm.nih.gov/geo/query/acc.cgi?acc=GSE153056}{GSE153056}. For our analysis we considered the top 2,000 highly variable genes returned from the Scanpy \texttt{highly\_variable\_genes} function with \texttt{flavor=seurat\_v3}. For the analysis presented in Section \ref{sec:results}, normalized protein counts were computed using the centered log ratio transform as implemented in muon \citep{bredikhin2022muon} with \texttt{axis=1} to match the original analysis of \citet{papalexi2021characterizing}. Based on the original analysis of \citet{papalexi2021characterizing}, cells with perturbations identified as having trivial effects were all labeled as nonperturbed and considered as control cells when training \method.

\textbf{\citet{norman2019exploring}.} The cell by gene count matrix for this dataset along with corresponding metadata was obtained from the NIH gene expression omnibus entry \href{https://www.ncbi.nlm.nih.gov/geo/query/acc.cgi?acc=GSE133344}{GSE133344}. As done in the analysis of \citet{norman2019exploring}, cells with the perturbation label \texttt{NegCtrl1\_NegCtrl0\_\_NegCtrl1\_NegCtrl0} were excluded from our analysis. Cells marked as doublets (i.e., a \texttt{number\_of\_cells} metadata value greater than 1.0) by \citet{norman2019exploring} were also excluded from our analysis. For our experiments we retained all cells with control guides along with cells infected with non-control guides that were annotated with gene program labels by \citet{norman2019exploring}. For our analysis we considered the top 2,000 highly variable genes returned from the Scanpy \texttt{highly\_variable\_genes} function with \texttt{flavor=seurat\_v3}.

\textbf{\citet{replogle2022mapping}.} For the analysis presented in this work we considered a filtered version of the original genome-wide data presented in \citet{replogle2022mapping} provided by \citet{bereket2024modelling} that retained data from perturbations with non-trivial effect sizes. Among this set of perturbations, for our experiments we considered the perturbations that had corresponding pathway labels provided by \citet{replogle2022mapping}, and we used the same set of highly variable genes considered in \citet{bereket2024modelling}.

\section{\method and baseline method implementation details}
\label{appendix:baselines}

In the experiments presented in Section \ref{sec:results}, we compared our proposed \method against two baselines: the original ContrastiveVI model of \citet{weinberger2023isolating} and the Mixscape method of \citet{papalexi2021characterizing}. Here we provide a brief overview of these methods and their corresponding implementations used in this work.

\subsection{\method}

Our implementation of \method was performed using the \texttt{scvi-tools} library \citep{gayoso2022python}. Our variational distributions $q_{\phi_z}$ and $q_{\phi_t}$ were implemented as multilayer perceptrons with a single hidden layer of 128 units and ReLU activation functions \citep{nair2010rectified}. For all experiments we set the dimensionality of both the background and salient latent spaces (i.e., $z$, and $t$) to 10. For $q_{\phi_y}$, we found that using additional hidden layers led to more consistent performance across random initialization, with good stability achieved with three hidden layers. Thus, for all results presented in this manuscript we implemented $q_{\phi_y}$ as an MLP with three hidden layers with 128 units each. For our decoder network we used an MLP with a single hidden layer of 128 units. All \method models were optimized using Adam \citep{kingma2014adam} with the default parameters in \texttt{scvi-tools}.

\subsection{ContrastiveVI}

The original ContrastiveVI model of \citet{weinberger2023isolating} extends the scVI model of \citet{lopez2018deep} via the contrastive latent variable modeling framework described in Section \ref{sec:background}. Specifically, for a cell $i$ labelled with a non-control gRNA, ContrastiveVI assumes the following generative process. Let
\begin{equation*}
    z_i \sim \mathcal{N}(0, I)
\end{equation*}
denote a low-dimensional set of \textit{background} latent variables capturing factors of variation found in both perturbed cells as well as controls. Next, let
\begin{equation*}
    t_i \sim \mathcal{N}(0, I)
\end{equation*}
denote a low-dimensional set of \textit{salient} latent variables capturing novel perturbation-induced variations in cells labeled with non-NTC guides.

Letting $f^{\eta}$ denote a neural network with a softmax non-linearity as the final layer, we then compute
\begin{equation*}
    \rho_{i} = f^{\eta}(z_i, t_i).
\end{equation*}
Analogous to scVI, this vector on the probability simplex represents the expected normalized expression frequency of each gene $g$. For a gene $g$ we then assume that the observed gene expression $x_{ig}$ in cell $i$ is drawn
\begin{equation*}
    x_{ig} \sim \text{ZINB}(\ell_i\rho_{ig}, \theta_{g}, f^{\nu}(z_i, t_i)),
\end{equation*}
where ZINB denotes the zero-inflated negative binomial distribution, $\ell_i$ is the observed library size for cell $i$, $\theta_g$ is a gene-specific inverse dispersion parameter, and $f^{\eta}$ is a neural network whose outputs are interpreted as dropout probabilities. For cells with NTC guides, ContrastiveVI assumes the same generative process but with the salient latent variables fixed at a constant zero vector. 

For inference, ContrastiveVI posits a variational distribution with parameters $\phi$ that factorizes as
\begin{equation*}
    q_{\phi}(z_i, t_i, \mid x_i) = q_{\phi_z}(z_i \mid x_i)q_{\phi_t}(t_i \mid x_i)
\end{equation*}
for cells with non-NTC guides. For cells with control guides, the above variational distribution is modified to account for the assumption that the salient variables do not contribute to the generative process, yielding
\begin{equation*}
    q_{\phi}(z_i, t_i, \mid x_i) = q_{\phi_z}(z_i \mid x_i)\delta\{t_i = 0\}.
\end{equation*}
In other words, the salient variables $t_n$ are simply fixed at 0 during inference for cells with control guides. We refer to \citet{weinberger2023isolating} for derivation of corresponding evidence lower bounds.

In our experiments we used the scverse \citep{virshup2023scverse} compatible implementation of ContrastiveVI available in the \texttt{scvi-tools} \citep{gayoso2022python} package. For our experiments we used the same model architecture and optimization hyperparameters as described in the ContrastiveVI paper \citep{weinberger2023isolating}.

\subsection{Mixscape}

The Mixscape procedure proposed in \citet{papalexi2021characterizing} begins by computing a so-called perturbation signature for each cell infected with a non-NTC guide. To do so, for a given cell with a non-NTC guide, that cell's nearest neighbors in the control population are identified. The gene expression profiles of these nearest control neighbors are then averaged together and substracted from the gene expression profile of the original given non-NTC guide cell. The result of this procedure is then defined as a cell's perturbation signature.

After computing cells' perturbation signature, Mixscape then classifies cells with non-NTC guides as perturbed or escaping perturbation. To do so, for each targeted gene Mixscape fits a mixture of Gaussian models with two components on the corresponding cells' perturbation signatures. As escaping cells' signatures are assumed to be similar to those of control cells, one component of each of these mixtures is constrained to be equal to of a unimodal Gaussian fit to control cells. 

For all results presented in this work, we used the R implementation of Mixscape in the Seurat package with default parameters. When computing UMAP embeddings and metrics on representation quality (e.g. mixing across cell cycle phases), we used the principal components of cells' perturbation signatures as done in \citet{papalexi2021characterizing}.

\section{Metrics}

Here we provide details on the quantitative metrics used in this work to compare \method against baseline methods.

\subsection{Entropy of mixing}
\label{appendix:entropy}

For $c$ groups (e.g. cell cycle phases.) the entropy of mixing \citep{haghverdi2018batch} is defined as

\begin{equation*}
    \sum_{i=1}^{c}p_{i}\log p_{i},
\end{equation*}

\noindent where $p_{i}$ denotes the proportion of cells from group $i$ in a given region, such that $\sum_{i=1}^{c}p_{i} = 1$. Next, let $U$ denote a uniform random variable over the population of cells. Let $B_U$ then denote the empirical proportions of cells' groups in the 50 nearest neighbors of cell $U$. We report the entropy of this variables averaged over 100 random cells $U$. Higher values of this metric indicate stronger mixing of the $c$ groups.

\subsection{Adjusted Rand index}
\label{appendix:ari}

The adjusted Rand index (ARI) measures agreement between reference clustering labels and labels assigned by a clustering algorithm. Given a set of $n$ samples and two sets of clustering labels describing those cells, the overlap between clustering labels can be described using a contingency table, where each entry indicates the number of cells in common between the two sets of labels. Mathematically, the ARI is calculated as

\begin{equation*}
    \text{ARI} = \frac{ \left. \sum_{ij} \binom{n_{ij}}{2} - \left[\sum_i \binom{a_i}{2} \sum_j \binom{b_j}{2}\right] \right/ \binom{n}{2} }{ \left. \frac{1}{2} \left[\sum_i \binom{a_i}{2} + \sum_j \binom{b_j}{2}\right] - \left[\sum_i \binom{a_i}{2} \sum_j \binom{b_j}{2}\right] \right/ \binom{n}{2}},
\end{equation*}

\noindent where $n_{ij}$ is the number of cells assigned to cluster $i$ based on the reference labels and cluster $j$ based on a clustering algorithm, $a_i$ is the number of cells assigned to cluster $i$ in the reference set, and $b_j$ is the number of cells assigned to cluster $j$ by the clustering algorithm. ARI values closer to 1 indicate stronger agreement between the reference labels and labels assigned by a clustering algorithm. In our experiments we used the $k$ means clustering algorithm to assign cluster labels to cells with $k$ equal to the true number of clusters.

\end{document}